\documentclass[12pt]{iopart}

\usepackage{graphicx}
\usepackage{xcolor}
\begin{document}

\title[Tachyon vs Quintessence]{Non-linear spherical collapse in tachyon models and a comparison of collapse in tachyon and quintessence models of dark energy}

\author{Manvendra Pratap Rajvanshi \& J.S. Bagla}

\address{Dept. of Physical Sciences, IISER Mohali, Sector 81, SAS Nagar, Punjab (India) - 14036}
\ead{\mailto{manvendra@iisermohali.ac.in}, \mailto{jasjeet@iisermohali.ac.in}}
\vspace{10pt}
\begin{indented}
\item[]September 2020
\end{indented}

\begin{abstract}
  We study evolution of perturbations in dark matter and dark
  energy for spherical collapse using a completely self consistent,
  relativistic approach.
  We study Tachyon models of dark energy using the approach outlined
  in Rajvanshi and Bagla (2018).
  We work with models that are allowed by current observations.
  We find that as with Quintessence models allowed by observations,
  dark energy perturbations do not affect evolution of perturbations
  in dark matter in a significant manner.
  Perturbations in dark energy remain small for such models.
  We then take two different Lagrangians for dark energy: tachyon and
  quintessence models, reconstruct potential to have same expansion
  history and then compare if two can be distinguished in the
  nonlinear regime. 
  Any variations we find are only due to a different Lagrangian
  density, and allow a comparison of different classes of models in a
  fair manner. 
  We find that dark matter perturbations carry no imprint of the class
  of dark energy models for the same expansion history: this is
  significant in that we can work with any convenient model to study
  clustering of dark matter. 
  We find that the evolution of dark energy perturbations carries an
  imprint of the class of models and dark energy perturbations grow
  differently in Tachyon models and Quintessence models for the same
  expansion history.
  However, the difference between these diminishes for $(1+w) \ll 1$
  and hence prospects for differentiating between models using
  characteristics of perturbations are limited in our Universe.
\end{abstract}

%
%
%
%
%

\section{Introduction}
\label{sec:intro}

Observations have shown that the Universe is undergoing accelerated
expansion \cite{Riess_1998,Perlmutter:1998np}.
This inference was the result of improvements in determination of
cosmological parameters through improved measurement of distances. 
In particular use of Supernovae type Ia as standardized candles
\cite{1993ApJ...413L.105P,1996AJ....112.2391H,Riess_1996,1995ApJ...438L..17R}
allowed distance determination to be made up to higher redshifts.
The observed accelerated expansion spurred developments in theoretical
cosmology as the {\sl obvious} explanation, the cosmological constant
\cite{1995Natur.377..600O,1996ComAp..18..275B} is riddled with fine
tuning and naturalness problems
\cite{RevModPhys.61.1,Bull:2015stt,Navarro:1999fr,DelPopolo:2016emo}.   
A number of approaches have been tried: one approach is where
Einstein's theory is modified in some manner
\cite{Clifton:2011jh,PhysRevLett.85.2236,Fujii_2003,Sotiriou:2008rp,Starobinsky:2007hu,Nojiri:2010wj,Nojiri:2017ncd}.
These changes affect the left hand side of Einstein's equation. 
This approach has to contend with the remarkable success of the
general theory of relativity when confronted by observational tests
\cite{Will:2014kxa,Yunes2013,PhysRevD.78.063503}. 
The other leading approach is the introduction of a constituent that
mimics the cosmological constant in an approximate manner.
This additional component, the so called dark energy, is required to
have some unusual properties
\cite{2010deto.book.....A,Buchert:2007ik,Huterer:2017buf,Copeland:2006wr}.
There are a large number of possibilities that have been proposed and
explored in this category.
This approach modifies the contents of the Universe and hence it
affects the right hand side of Einstein's equations. 
Each category corresponds to a specific action for a component, in
most such models the dark energy couples minimally with other
constituents of the Universe. 
Some examples of such models are scalar fields, K-essence, tachyon
models, Chaplygin gas,
etc. \cite{Ratra:1987rm,PhysRevD.67.063504,ArmendarizPicon:2000dh,ArmendarizPicon:2000ah,2001PhLB..511..265K},
(see \cite{2010deto.book.....A} for review). 
In each of these models the dynamics of the Universe approximates the  
cosmological constant in order to reproduce accelerated expansion.
Present observations allow for small deviations from the cosmological
constant model. 
Thus, there are qualitative and quantitative differences in the
dynamics, though each model can be tuned to produce the expansion
history required by observations within some reasonable constraints.  

A fundamental difference between the cosmological constant and other
models is that the cosmological constant does not vary with time or
location, whereas other dark energy models allow for such variations. 
In all other models the dark energy component is allowed to vary and
respond to variations in the gravitational field.
A number of studies have been carried out to study dynamics and
perturbations in various dark energy models
\cite{Doran:2003xq,Malquarti:2002iu,Abramo:2001mv,Unnikrishnan:2008qe,Jassal:2009ya,PhysRevD.66.021301,PhysRevD.67.063504,2019JCAP...04..047S,Singh:2019bfd}. 
The key result of these studies, obtained using linear perturbation
theory or other approximations, is that the perturbations in dark
energy remain very small.
However, perturbation theory is valid only at early times or at very
large scales at late times.
Thus it cannot be used to study dark energy perturbations and their
interplay with highly non-linear dark matter perturbations at small
scales. 

In an earlier work we have studied fully non-linear evolution of
spherically symmetric perturbations in quintessence models of dark
energy \cite{2018JCAP...06..018P,erratum}.
We found that the amplitude of dark energy perturbations remains small
in all cases.
We also found that the effective equation of state parameter of dark
energy becomes a function of coordinates and this variation is
correlated with the density contrast of dark matter.  

Here we use the same methodology and study tachyon models for dark
energy.\\
There are low energy effective theories that arise
  from string theory that contain tachyon fields \cite{Sen:2002in}
  with Lagrangian:
    \begin{equation}
    \mathcal{L} = -V(\psi)\sqrt{(1-\partial^\mu\psi \partial_\mu\psi)}
  \end{equation}
  Here $\psi$ is tachyon field and $V(\psi)$ is potential.
  As an analogy, if one sees quintessence a field form of classical
  particle Lagrangian(kinetic term+ potential part), then tachyon
  Lagrangian is field form of Lagrangian for relativistic particle.  
  Tachyon models and their characteristics have been studied in
  detail\cite{PhysRevD.66.021301,PhysRevD.67.063504}.
  As shown in \cite{PhysRevD.67.063504}, some potentials (particularly
  exponential potential $V\propto e^{-\psi}$) have interesting
  asymptotic future behavior with the possibility to avoid future
  horizon.
  There have also been some attempts to unify dark matter and dark
  energy in terms of a single tachyon field \cite{Padmanabhan:2002sh}.
  Here, inverse square potential($V\propto \psi^{-2}$) as a function
  of field $\psi$ averaged over some scale gives a dark matter like
  behaviour at certain scales.
  While quintessence models have been extensively studied in context
  of various types of perturbations, tachyonic models have not been
  studied in as much detail (see \cite{Singh:2019bfd} for study of
  linear perturbations in tachyon models).  
  Different theoretical motivations/insights might lead to different
  class of models, but there has to be framework that can be
  used to distinguish different type of models.
  It is in this context, that we carry on from our previous
  work\cite{2018JCAP...06..018P} where we simulated spherical collapse
  for quintessence, modify the formalism for tachyonic field and do a
  systematic comparison.    
We study two potentials( $V\propto \psi^{-2}$ and $V\propto
e^{-\psi}$) that have been proposed and studied for tachyon
models because of their interesting features as discussed above (see
\cite{PhysRevD.67.063504,PhysRevD.66.021301}). 
Further, in order to explore the dependence of the growth of
perturbations on the class of models, we compare the evolution of
perturbations in quintessence models and tachyon models \textit{for
  the same expansion history}. 

We describe the formalism and equations in \S{\ref{sec:equations}}.
Details of the
expansion history in models to be studied is discussed in
\S{\ref{sec:results_for_back}} for two potentials studied here for
tachyon models.  Evolution of perturbations for dark matter and dark
energy in these cases is described in \S{\ref{sec:perturbations}}.
We then proceed to compare quintessence and tachyon models by working
with potentials that give us the same expansion history.
These are discussed in \S{\ref{sec:results_compare}}.
Results are summarised in \S{\ref{subsec:dm_results}} and
\S{\ref{subsec:de_results}}, dealing with dark matter properties and
dark energy properties respectively. 

\section{Equations and Formalism}
\label{sec:equations}

We follow the scheme set out in Rajvanshi and Bagla
\cite{2018JCAP...06..018P,erratum} and refer the reader to the paper
for more details. 

We assume spherical symmetry and treat dark matter as a pressure-less
fluid.
Tachyon models are described by the following Lagrangian density:
\begin{equation}
\mathcal{L} = -V(\psi)\sqrt{(1-\partial^\mu\psi \partial_\mu\psi)}
\label{eq:tachyon_lagrange}
\end{equation}
Space-time is described by the following metric:
\begin{equation}
    \label{eq:1} 
    ds^2\,=\,-e^{(2B)}dr^2 - R^2(d\theta^2 + sin^2\theta d\phi^2) +
    dt^2      
\end{equation}
where $B(r,t)$ and $R(r,t)$ are unknown functions of comoving radial
coordinate $r$ and time $t$.

These allow us to obtain dynamical equations for all the variables in
the system.
 
  Note: We work in units where speed of light $c$ and gravitational
  constant $G$, both are unity.
  $V$ denotes potential as a function of $\psi$ and $V_{,\psi}$
  represents gradient of this potential with respect to $\psi$. 
The full set of equations along with Einstein's equations is:
\begin{eqnarray}
\ddot{B} &=& -e^{-2B}\frac{R'^2}{R^2} + \frac{1}{R^2} +
\frac{\dot{R}^2}{R^2} - \dot{B}^2 - 4\pi \rho_{dm} + 4\pi V
\left[ \frac{e^{-2B}\psi'^2 - \dot{\psi}^2}{\sqrt{1-u^2}} \right]  
\label{eq:Q26} \\
\frac{\ddot{R}}{R} &=& \frac{4\pi V}{\sqrt{1-u^2}}\left[
  1-u^2 -  e^{-2B}\psi'^2 \right]
-\frac{1}{2}\frac{\dot{R}^2}{R^2} + \frac{1}{2} \left[ 
  e^{-2B}\frac{R'^2}{R^2} - \frac{1}{R^2} \right]   \label{eq:Q27}
\end{eqnarray}
\begin{eqnarray}
\ddot{\psi}RV(e^{2B}+\psi'^2) &=& 2e^{-2B}VR'\psi'^3
-2V\dot{R}\dot{\psi}\psi'^2 + 2VR'\psi'(1-\dot{\psi}^2) -RV_{,\psi}
\psi'^2\nonumber\\
&&- RVB'\psi'(1-\dot{\psi}^2) +
RV\psi''(1-\dot{\psi}^2)-2RV\dot{\psi}\dot{B}\psi'^2\nonumber\\
&&+2RV\dot{\psi}\psi'\dot{\psi}' 
-RV_{,\psi}e^{2B}(1-\dot{\psi}^2)\nonumber\\
&&-V\dot{\psi}(1-\dot{\psi}^2)(R\dot{B}+2\dot{R})e^{2B}\label{eq:Q14}\\
\dot{\rho_{dm}} =-\left( \dot{B} + \frac{2\dot{R}}{R} \right)\rho_{dm}
\label{eq:Q25}
\end{eqnarray}
where $u^2 =\partial^\mu\psi \partial_\mu\psi $ and $\rho_{dm}$ represents dark matter density.\\
Here a prime represents a partial derivative with respect to
$r$ and a dot represents a partial derivative with respect to $t$.

We study evolution of perturbations for two potentials with tachyon
models, details of the potentials are given in the following
discussion.

In order to compare evolution of perturbations in tachyon models with
quintessence models, we work with potentials that lead to the same
expansion history.
Methods for computing the potential given an expansion history have
been developed for a variety of models
\cite{Chevallier:2000qy,Linder:2002et}. 
This process is often called reconstruction of potentials.
We have done these calculations for $w=$~constant and Chevallier-Polarski-Linder(abbreviated as "CPL" from here on, see equation \ref{eqn:cpl}) parametrization for quintessence and tachyon models, details of the
approach are given in \cite{Rajvanshi:2019wmw}. 
\subsection{Computational Methods}
  We consider a 1-d discrete grid in radial variable $r$, and the
  dependent variables (fields and their first time derivatives) are
  simulated on this grid as a function of $r$ which are evolved in
  time using fourth order Runge-Kutta  scheme(RK-4). At each time
  instant $t_i$ we calculate all spatial derivatives using finite
  difference schemes, this allows us to write all first order time
  derivatives(including derivatives of 1st time derivatives
  i.e. accelerations) of dependent variables as functions of
  quantities at $t_i$. These functions allow us to get prediction for
  1st sub-step of RK-4 scheme and temporary values of all dependent
  variables which are used for calculations of further sub-steps. This
  process is repeated until the time for the final intended output is
  reached.
  We check for numerical stability and convergence by running for
  different time steps.
  The computational methodology is described in detail in paper
  I\cite{2018JCAP...06..018P}.

\section{Results: Background Evolution}
\label{sec:results_for_back}

\begin{figure}[hbt!]
    \centering
    \includegraphics[width=1.1\textwidth]{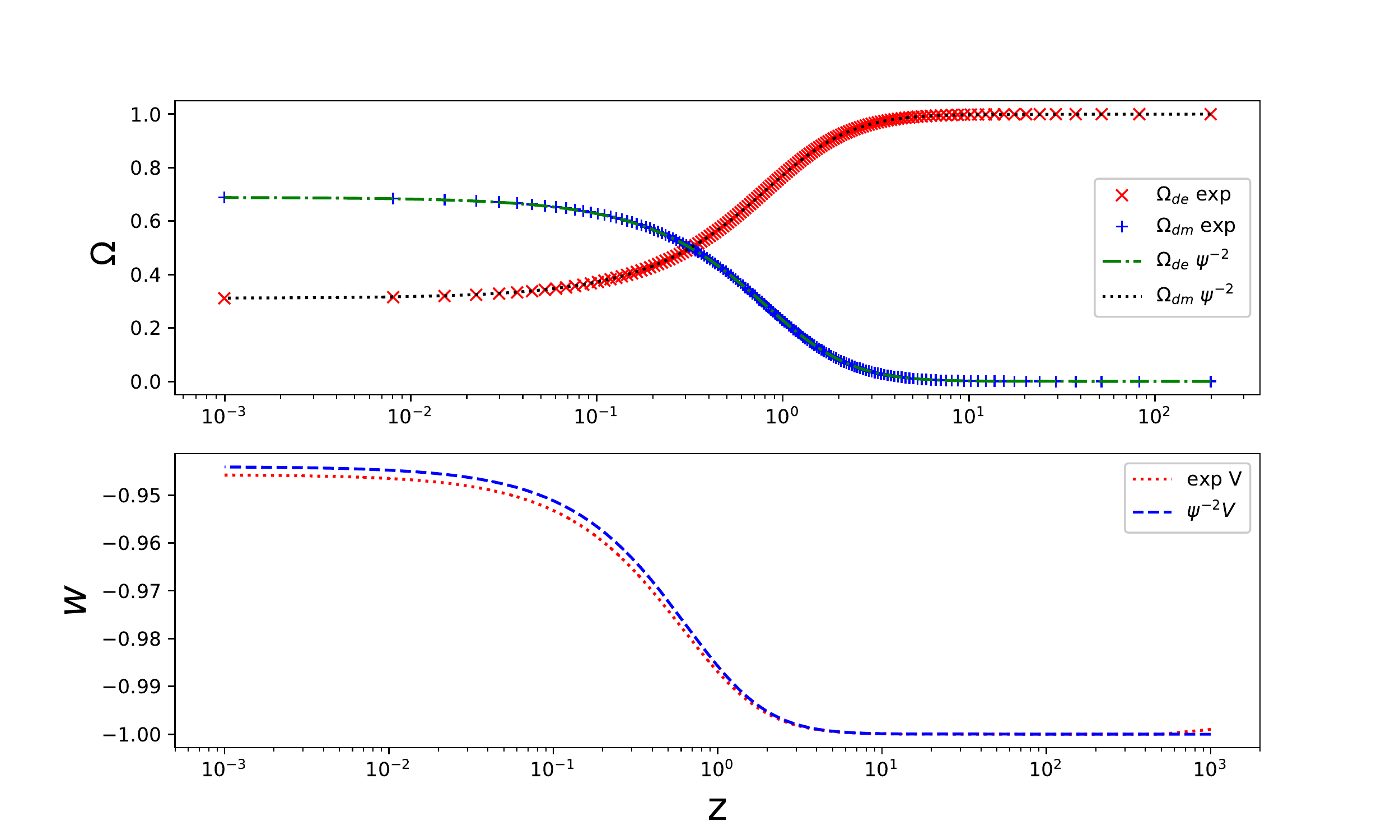}
    \caption{Energy densities(Upper Panel) contribution of dark matter
      and dark energy as a function of redshift($z$). Bottom panel
      show the evolution of equation of state($w$) of tachyon
      field. Both backgrounds are very similar in terms of
      observations with slight difference in effective equation of
      state parameter($w$).} 
    \label{fig:back}
\end{figure}

We use two potentials for tachyon models that have been studied
extensively.
We study the background evolution for potentials $V\propto
\psi^{-2}$ and $e^{-\psi}$.
Figure \ref{fig:back} shows the evolution of density parameters for
the tachyon field and dark matter, and the equation of state parameter
($w$).
Although each of these potentials has a unique asymptotic
behaviour\cite{PhysRevD.67.063504}, here we have tuned the
parameters such that they satisfy observational constraints
\cite{2019JCAP...04..047S}. 
Both the models shown here have a thawing behaviour.
These plots illustrate the generic behaviour in tachyon models that is
consistent with observations.
More details for background evolution and comparison with observations
can be found in the detailed study by Singh et. al \cite{2019JCAP...04..047S}. 

\section{Evolution of perturbations}
\label{sec:perturbations}

\begin{figure}[hbt!]
  
    \begin{minipage}{0.48\textwidth}
    \includegraphics[width=1.2\textwidth]{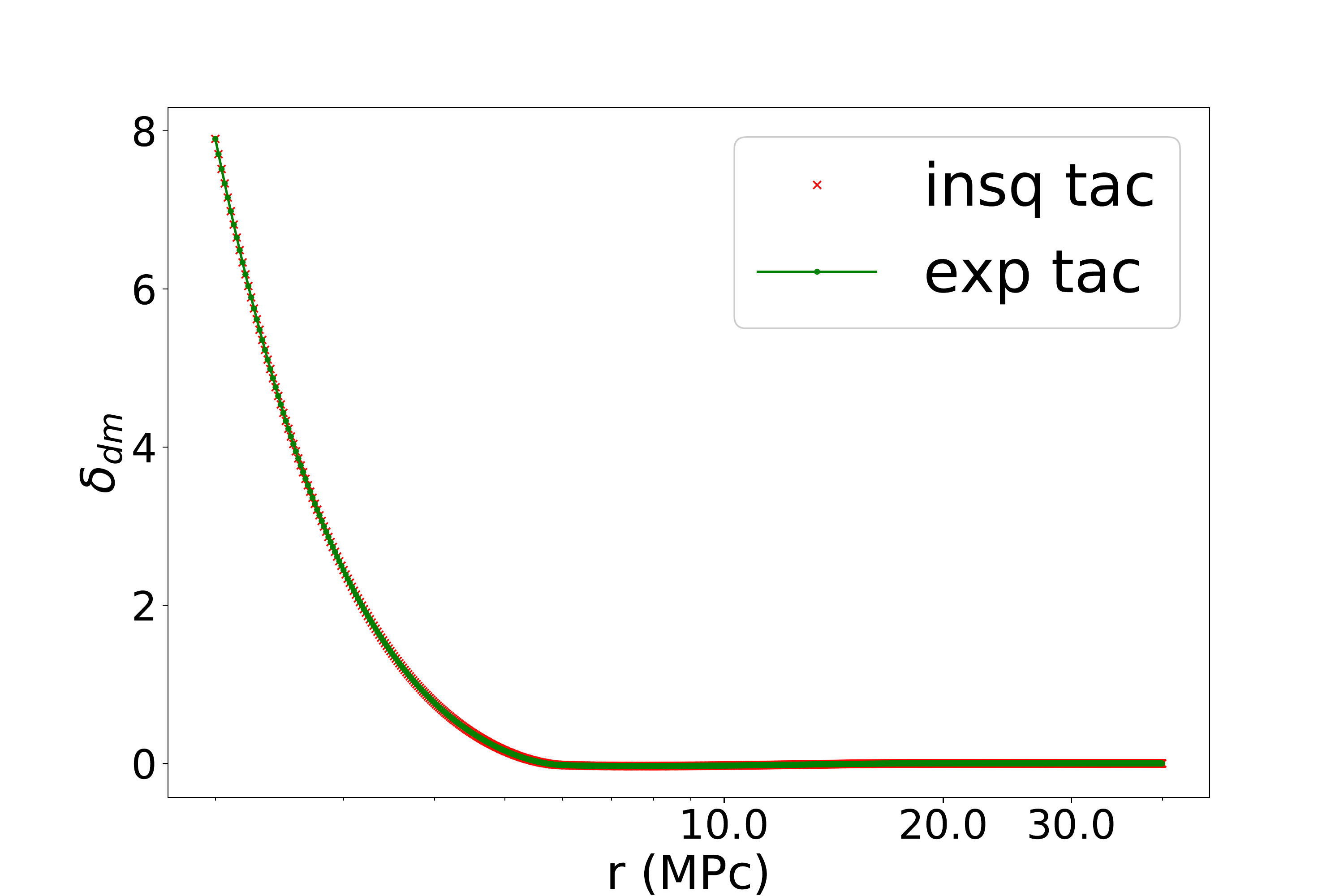}
    \caption{ Dark matter density contrast as a function of comoving
      radius at $z\sim 1.5$. The two curves correspond to two
      different dark energy potentials.
      Label 'insq' refers to $V\propto \phi^{-2}$ and 'exp' refers to
      $V\propto \exp$.}  
    \label{fig:delta_dm_2comp_odtach}
    \end{minipage}\hfill
     \begin{minipage}{0.48\textwidth}
    \includegraphics[width=1.2\textwidth]{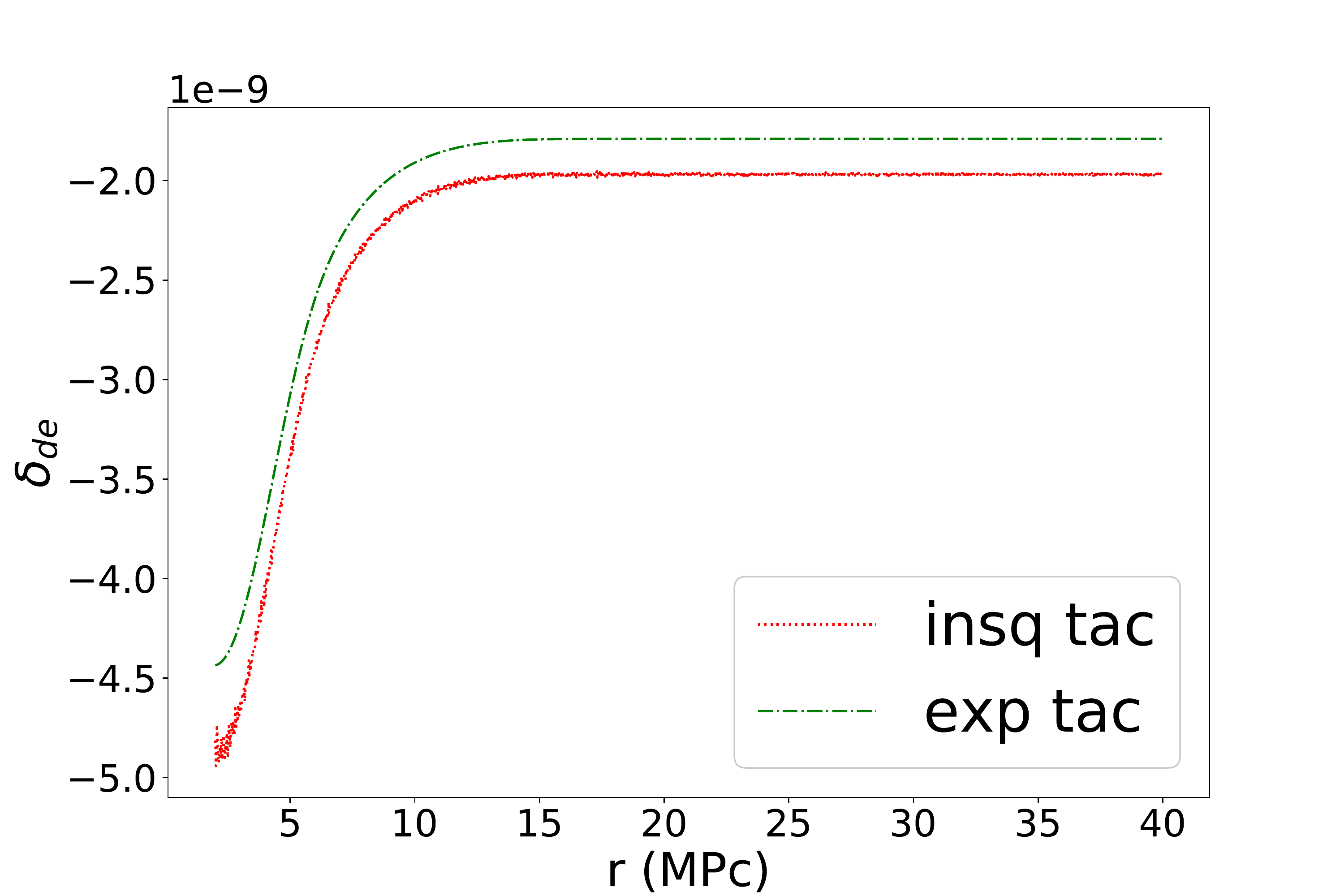}
    \caption{Dark energy density contrast as a function of comoving
      radius at $z\sim 1.5$.  Label 'insq' refers to $V\propto
      \phi^{-2}$ and 'exp' refers to $V\propto \exp$. } 
    \label{fig:delta_de_2comp_odtach}
    \end{minipage}
\end{figure}

We study perturbations in dark matter and dark energy for two
potentials: the exponential potential and the inverse square
potential. 
The initial conditions are set such that the dark matter does not have 
any peculiar velocities at the initial time.
Dark matter has an initial density perturbation. This
  initial density perturbation has a compensated profile
  i.e. $\delta_{dm}$ integrated from center to outermost radius comes
  out to be zero, so that average density contrast is 0. Please see
  \cite{2018JCAP...06..018P} for details of initial profile. 
Dark energy is set to have no perturbations at the initial time.
We find that such an initial condition quickly leads to the expected
adiabatic mode at early times. 
We start at $z_{ini} = 10^3$ and evolve the system towards lower
redshifts.  
We first study the evolution of an over-density.
A note for figures: In figures we often use scientific notation for quoting numbers i.e. format $a$ x $10^{b}$ with exponent part quoted on top of figure, so any no. that is quoted on top left of figure has to be multiplied to y-axis values to get actual values. Exponents are in powers of 10. For example, in figure \ref{fig:delta_de_2comp_odtach}, one has to multiply $10^{-9}$ to y values.
Figure~\ref{fig:delta_dm_2comp_odtach} is a plot of dark matter
density contrast at the time when
the inner regions begin to virialize at $z \simeq 1.5$.

The overdensity has been evolved from
$z\sim 1000$. This is for an initially overdense(OD) system.
This is shown for the two potentials we are studying and it can be
seen that the dark matter density contrast in these two cases is
indistinguishable. 
This similarity results from an almost identical expansion history. 
  
Corresponding plots for density contrast in dark energy are shown in
Figure~\ref{fig:delta_de_2comp_odtach}. This is for
  initially overdense (OD) system.
  At the start of simulation($z\sim 1000$) there was no perturbation
  in dark energy field, but metric perturbations induce perturbations
  in dark energy sector which grow with time. 
We see that the perturbation in dark energy is very small for the two
cases, but there are apparent differences in the two curves.

\begin{figure}[hbt!]
    \centering
    \includegraphics[width=0.7\textwidth]{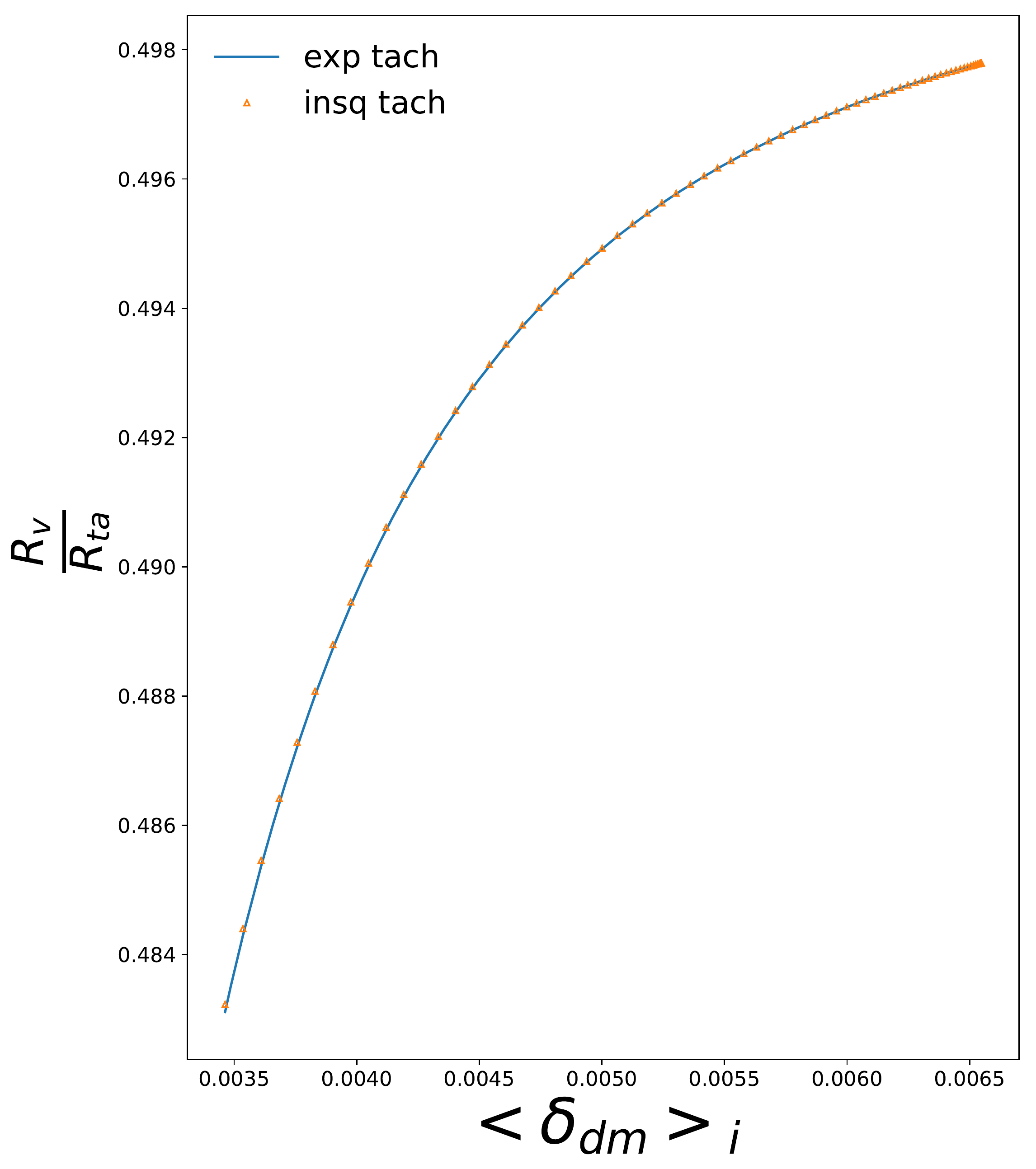}
    \caption{Ratio of virial radius to turn around(maximum) radius as
      a function of initial matter overdensity averaged over interior
      $r$ till that particular $r$. The ratio tends toward
      Einstein-DiSitter value of $0.5$ as the initial overdensity
      tends to infinity i.e. dark energy effects on perturbation
      become less significant as dark matter perturbation become
      stronger.} 
    \label{fig:b2sqr}
\end{figure}

The dark matter perturbation reaches maximum radius, called the turn
around radius, before collapsing back and eventually reaching
dynamical or virial equilibrium.
The ratio of virial to turn around radius is plotted in
Figure~\ref{fig:b2sqr}.
It is apparent from this plot that there is no discernable difference
in the values for the two potentials mainly because of a similar
expansion history. 

We now turn our attention to evolution of an under-dense region.
While an under-dense region is limited to $\delta_{dm} \geq -1$,
whereas the density contrast for an over-dense region can be very
large.
On the other hand, a realistic over-dense region with a large density
contrast cannot be arbitrarily large in size, whereas underdense
regions can easily be tens of Mpc across.
In terms of analysis, we also avoid loosing information inside the
virialized region as the equations cannot be solved self consistently
in this region \cite{2018JCAP...06..018P,erratum}. 

\begin{figure}[hbt!]
    \centering
    \includegraphics[width=1.2\textwidth]{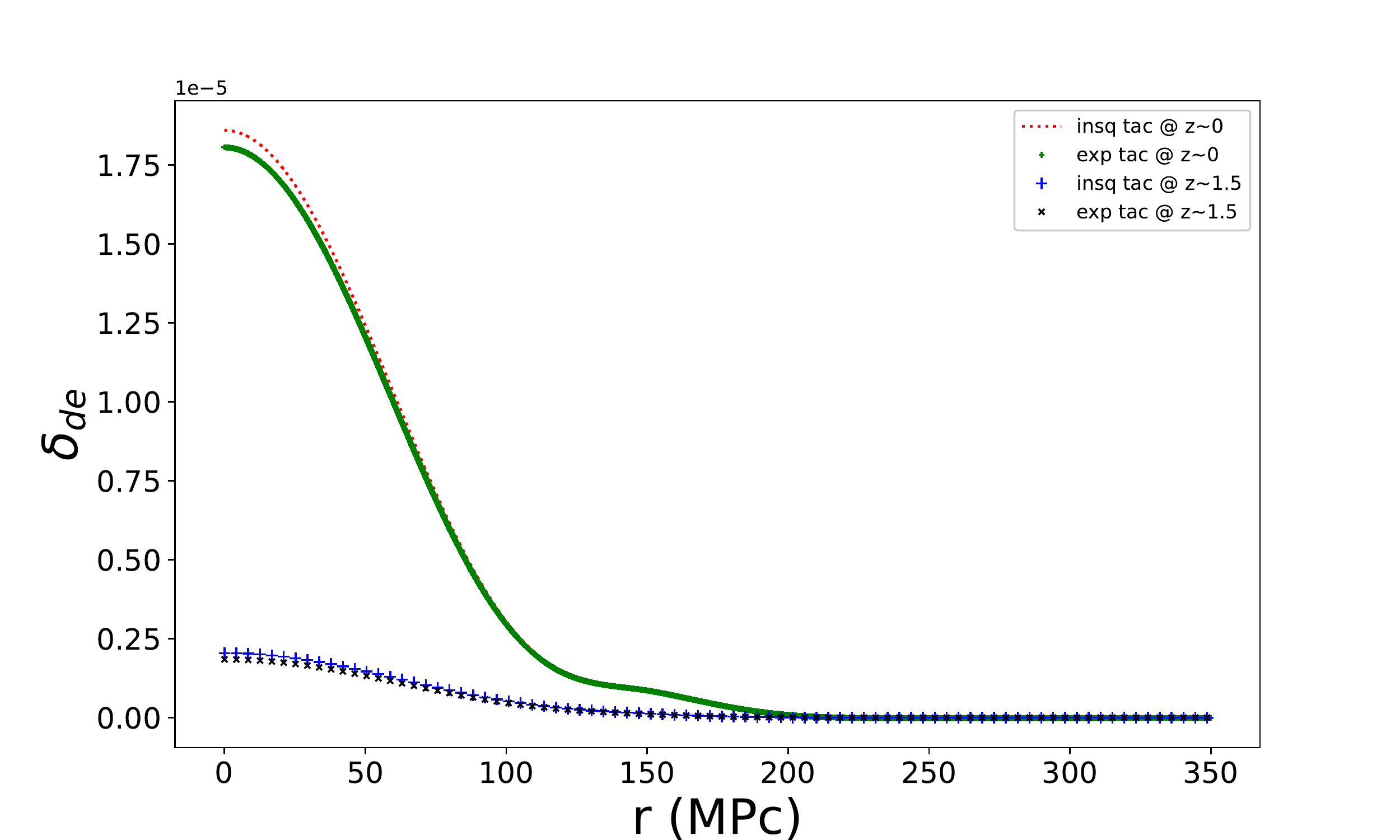}
    \caption{Dark energy(DE) density contrast as a function of comoving
      radius at two different redshifts. Here the initial matter
      perturbation was underdense. There was no perturbation in DE at
      initial time, but metric perturbations induce perturbation in DE
      field. This perturbation grows stronger with time as can be seen
      from curves at 2 different redshifts.}  
    \label{fig:delta_de_2comp_udtach}
\end{figure}

The dark energy perturbations are shown in
Figure~\ref{fig:delta_de_2comp_udtach}.
The density contrast is significant over the scale of the under-dense
region.
We also observe a rapid growth of dark energy perturbations at late
times, even though the amplitude of perturbations remains small at all
times.
We see some variation between the two potentials but it remains at a
few percent level and this can be attributed to the difference in
expansion history.

We note that the qualitative behaviour of perturbations in dark matter
and dark energy closely follows that seen for quintessence models
studied earlier \cite{2018JCAP...06..018P,erratum}.
In the following discussion we focus on a comparison of quintessence
and tachyon models for the same expansion history. 

\section{Results: Evolution of Perturbations in Quintessence vs
  Tachyon Models}
\label{sec:results_compare}

In order to compare perturbations in tachyon vs quintessence models of
dark energy, we reconstruct potentials in both models that correspond
to the same expansion history.
We codify the expansion history by the variation of the equation of
state parameter for dark energy with the scale factor $w(a)$.
Details of the procedure adopted for computing the potential are given in
\cite{Rajvanshi:2019wmw}.
We work with two different forms for $w$ for this comparison: $w =
constant$ and CPL \cite{Chevallier:2000qy,Linder:2002et}. 
We choose three values of constant $w$ for comparison and numerically
reconstruct the corresponding potentials for quintessence and
tachyonic fields,  
\begin{eqnarray}
    w = -0.5,\quad w = -0.9\quad and\quad w = -0.975 
\end{eqnarray}
and for CPL parametrization\cite{Chevallier:2000qy,Linder:2002et}
with form $w(a)$: 
\begin{eqnarray}
    w = w_0+w_a(1-\frac{a}{a_0})
    \label{eqn:cpl}
\end{eqnarray}
we have $w_0 = -0.9$ and $w_a = \pm 0.09$.
That is, the present day equation of state parameter is $-0.9$ in both
the cases but in one case it decreases as we go to earlier epochs, and
in the other it increases as we go to earlier epochs. In figures we represent cases with
$w_a = + 0.09$ with notation "cpl+" and $w_a = - 0.09$ model with "cpl-".
We investigate turn around and virialization characteristics for overdense
regions for these cases.  

\subsection{Dark Matter Perturbations}
\label{subsec:dm_results}

\begin{figure}[hbt!]
\centering 
\includegraphics[width=0.8\textwidth]{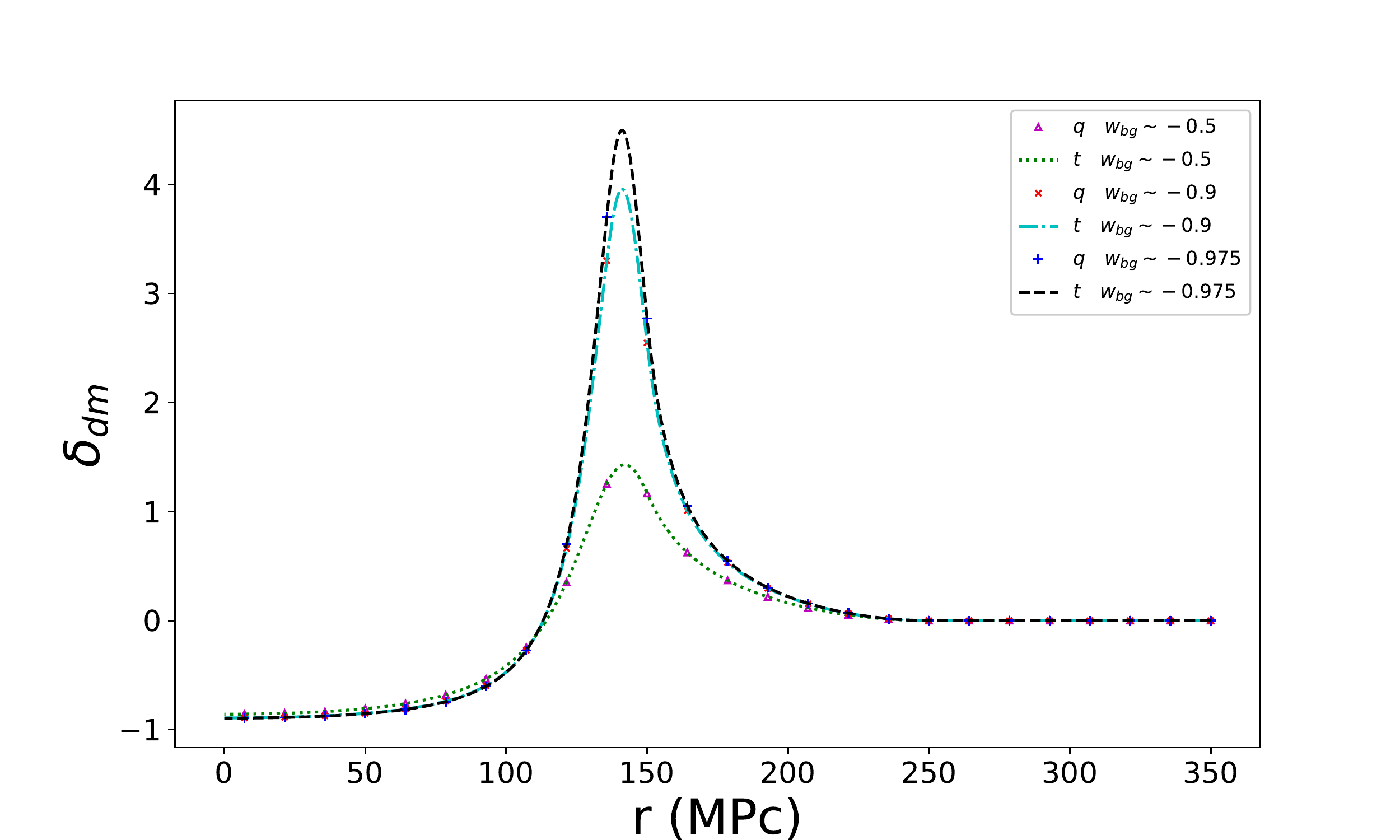}
\caption{\label{fig:3}Underdense cases constant w comparison: Dark
  matter density contrast evolved to $z\sim 0$. "q" refers to
  quintessence models while "t" for tachyonic models. "$w_{bg}$" is constant value of background 
  $w$. Curves are
  clustered by background histories with quintessence and tachyonic
  models with same background having indistinguishable matter
  perturbation dynamics.}      
\end{figure}

\begin{figure}[hbt!]
\centering 
\includegraphics[width=0.8\textwidth]{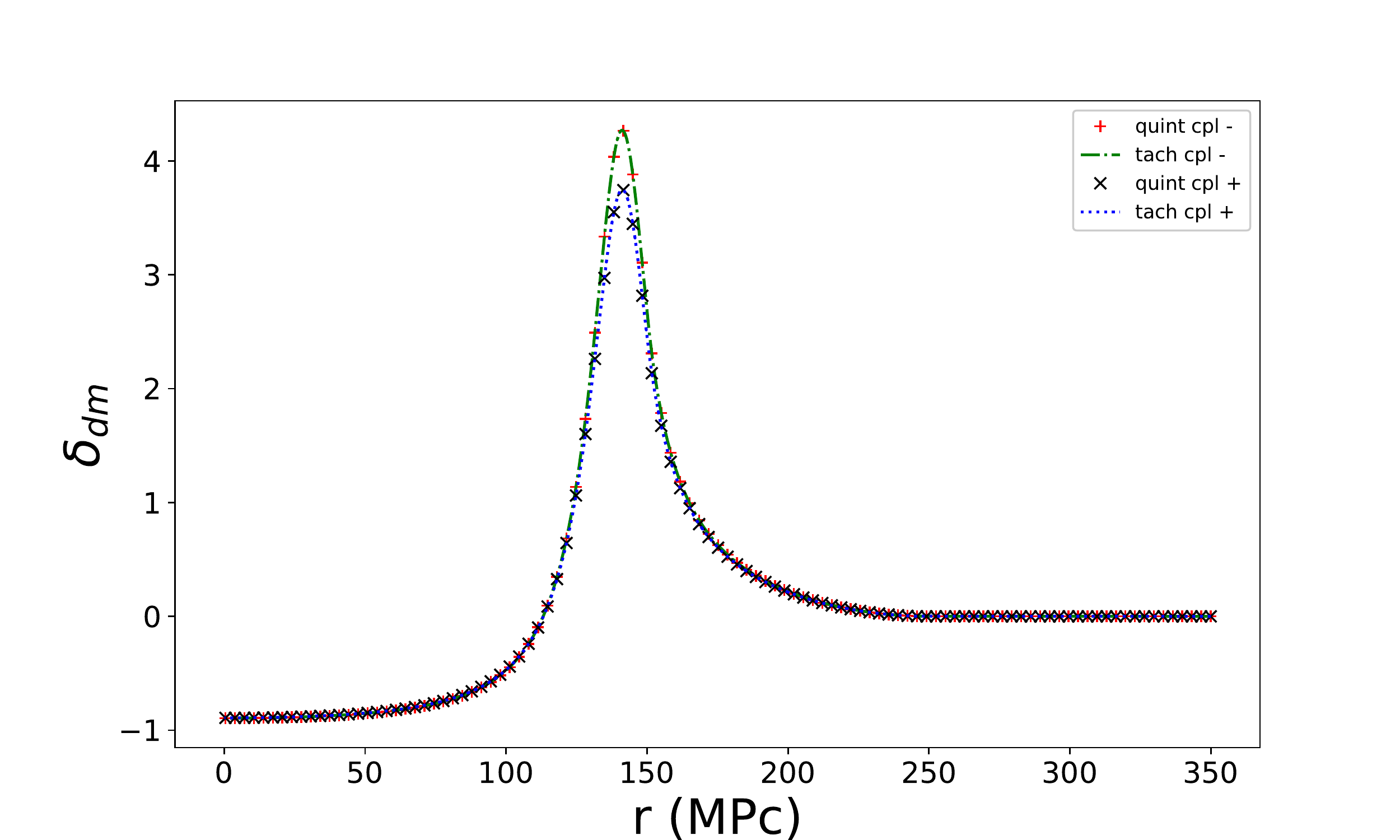}
\caption{\label{fig:3cpl}Underdense cases CPL:Dark matter density
  contrast evolved to $z\sim 0$ . cpl+ denotes $w_a = +0.09$ case and cpl-
  represents  $w_a = -0.09$. "quint" stands for quintessence and
  "tach" stands for tachyonic. As with constant $w$ cases, it is
  background evolution that is distinguishing the models rather than
  field dynamics Lagrangian being quintessence or tachyonic type.}      
\end{figure}

We have run our simulations setting initial conditions in the early
universe (at $z\sim 1000$) for underdense and overdense dark matter
perturbations.
We start with an unperturbed dark energy (see
\cite{2018JCAP...06..018P,erratum} for details of initial conditions).
The density contrast at present time is shown in figure~\ref{fig:3}
for constant $w$ for underdense initial condition.
We see that the density contrast for different expansion histories
differs from each other but {\sl there is no difference in the profile
  for quintessence and tachyon models.}
This clearly implies that the choice of dark energy model (tachyon or
quintessence) has no discernable impact on dark matter density
profiles in an underdense region as long as the expansion history is
the same.
Next we proceed to study the same in the two cases for the CPL
parameterization.
We refer to models by the sign of the term $w_a$ and $w_0$ is same in
the two cases ($w_0=-0.9$).
The two cases differ as we have $w_a=\pm 0.09$.
We show the dark matter density profile for the same initial condition
as above in Figure~\ref{fig:3cpl}.
Again, we find that there are distinctions between the two cases with
a different expansion history but there is no discernable difference
in the dark matter density profile for the two different models of
dark energy.
This is remarkable.
   Note that bump in contrast around $150$~Mpc is
   because of the compensating overdense region at edge of void to
   ensure that we go over to an FLRW universe at large $r$.

\begin{figure}[hbt!]
\centering 
\includegraphics[width=.45\textwidth]{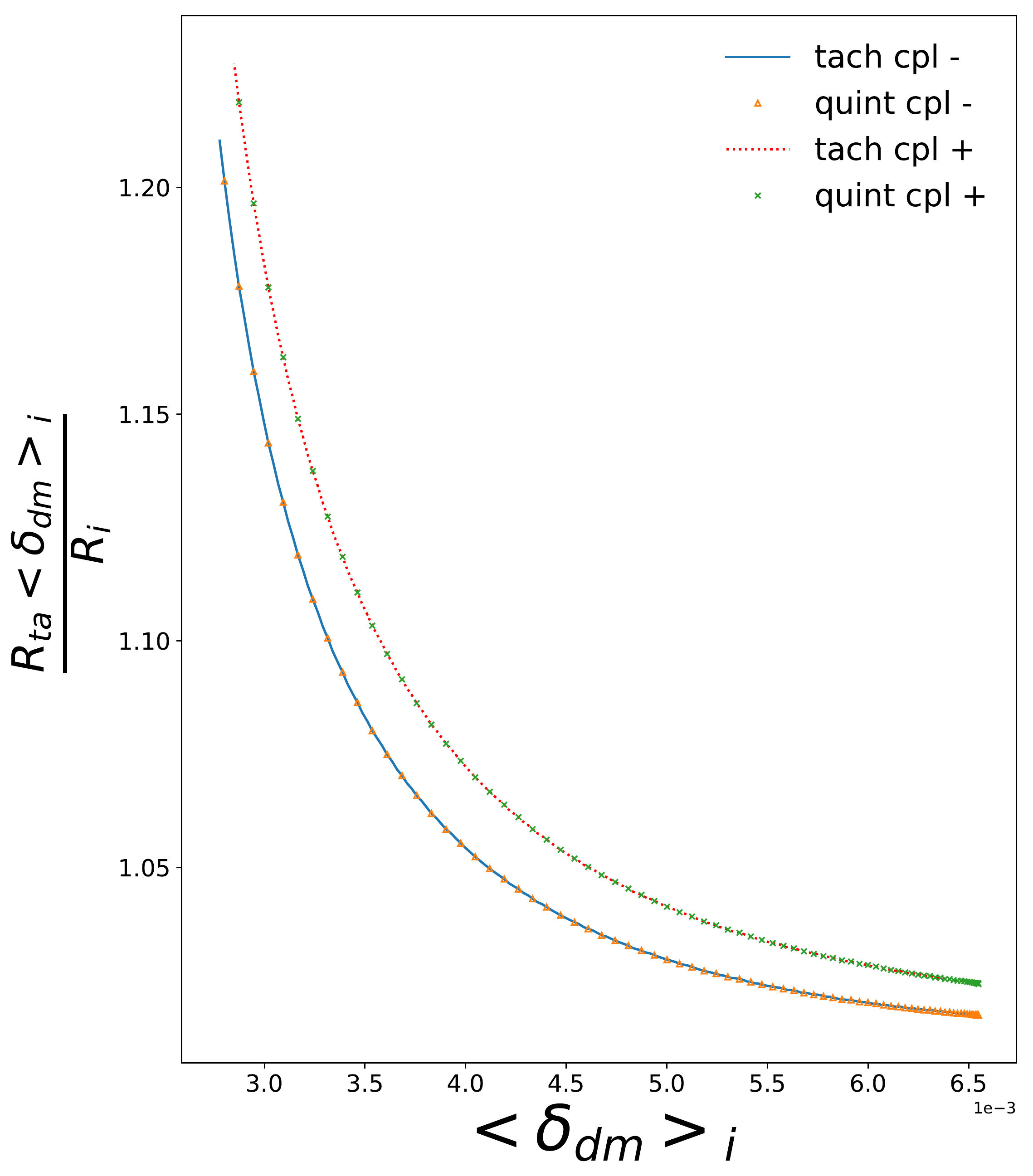}
\includegraphics[width=.42\textwidth]{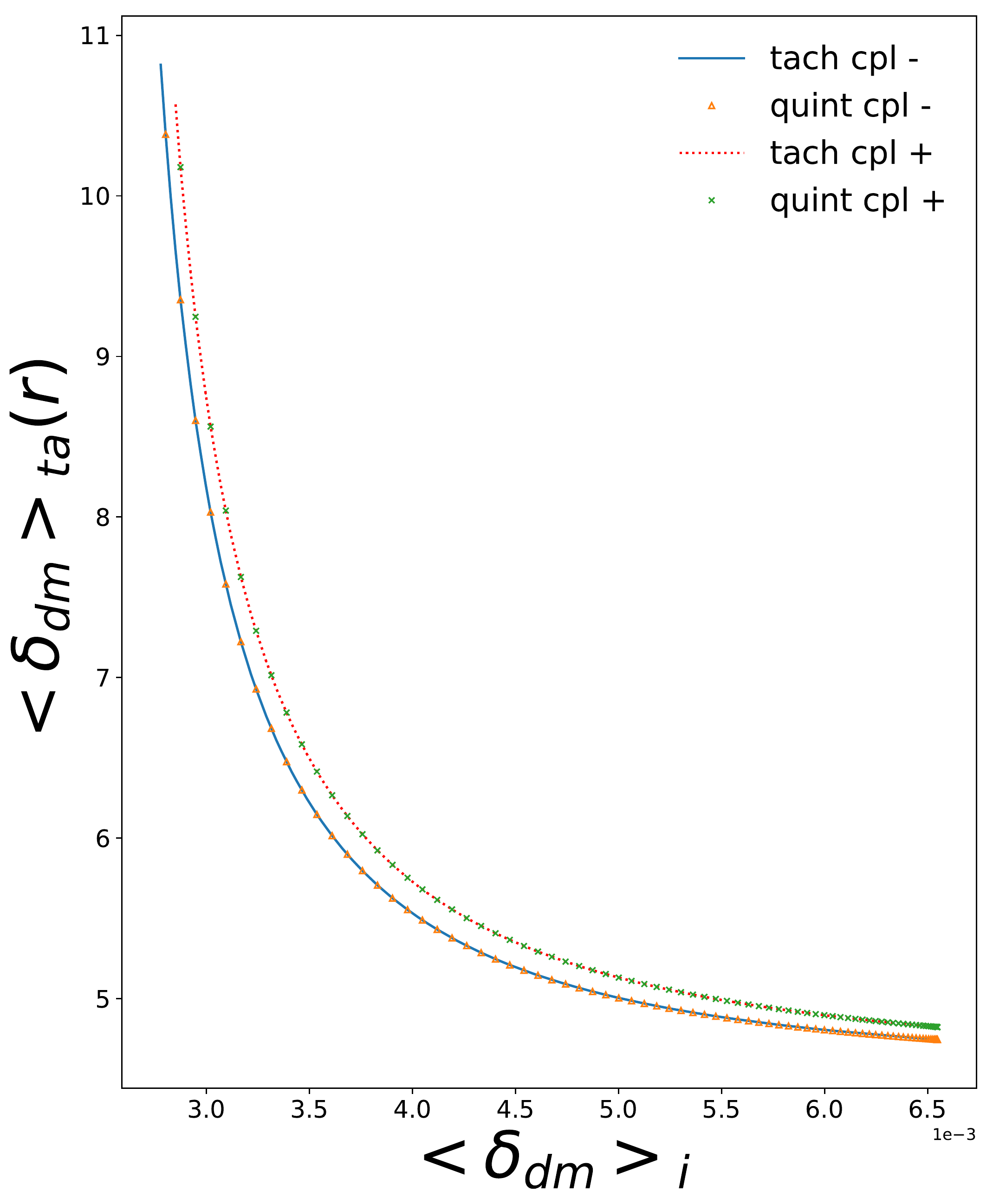}
\caption{\label{fig:5cpl}
 Turn around characteristics for CPL case. Left panel shows turn
 around radius in the combination 
  $R_{ta}\langle \delta_{dm} \rangle_i / R_i$ as a function of the
  initial density contrast. Right panel shows density contrast at turn around
  as a function of the initial density contrast. quint denotes
  quintessence and tach represents tachyonic field. cpl+ denotes $w_a =
  +0.09$ and cpl- represents  $w_a = -0.09$. }
\end{figure}

We now turn our attention to growth of overdensities in dark matter.
In these cases the perturbations collapse to form virialized halos if
the initial density is higher than a critical value as in the case for
$\Lambda$CDM \cite{1993MNRAS.262..717B}. 
Results for the two CPL parameterizations are shown here.
We show the characteristics of perturbations at turn around in
Figure~\ref{fig:5cpl}.
Variation of the turn around radius as compared to the expected value
in the Einstein-deSitter model as a function of the initial
overdensity is shown in the left panel. 
The right panel shows the density contrast at turn around as a
function of the initial dark matter overdensity.
Note that the overdensities are always volume averaged, so as to
facilitate comparison with the Einstein-deSitter and the $\Lambda$CDM
models.
The qualitative behaviour seen in the two panels is very similar to
what is known for the $\Lambda$CDM model in that the turn around
radius becomes very large as we approach the critical initial
overdensity from above.
The density contrast also increases in this limit as the time taken to
reach turn around increases and the background also increases and the
average density of the universe decreases to give us an enhanced
density contrast.
The two CPL models representing two different expansion histories lead
to different curves.
However, there is no obvious difference between the tachyon and
quintessence models for a given expansion history.

\begin{figure}[hbt!]
\centering 
\includegraphics[width=.45\textwidth]{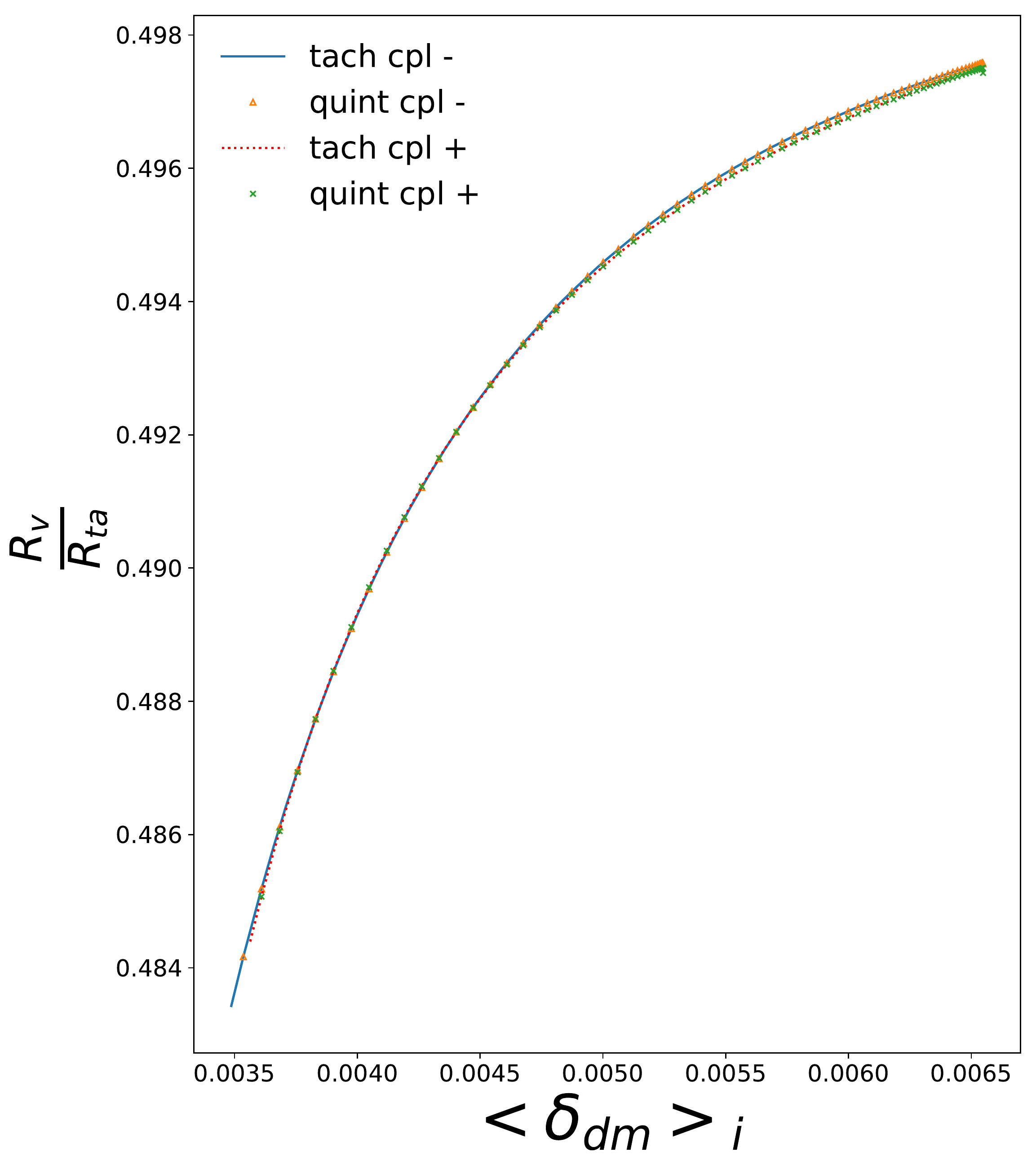}
\includegraphics[width=.45\textwidth]{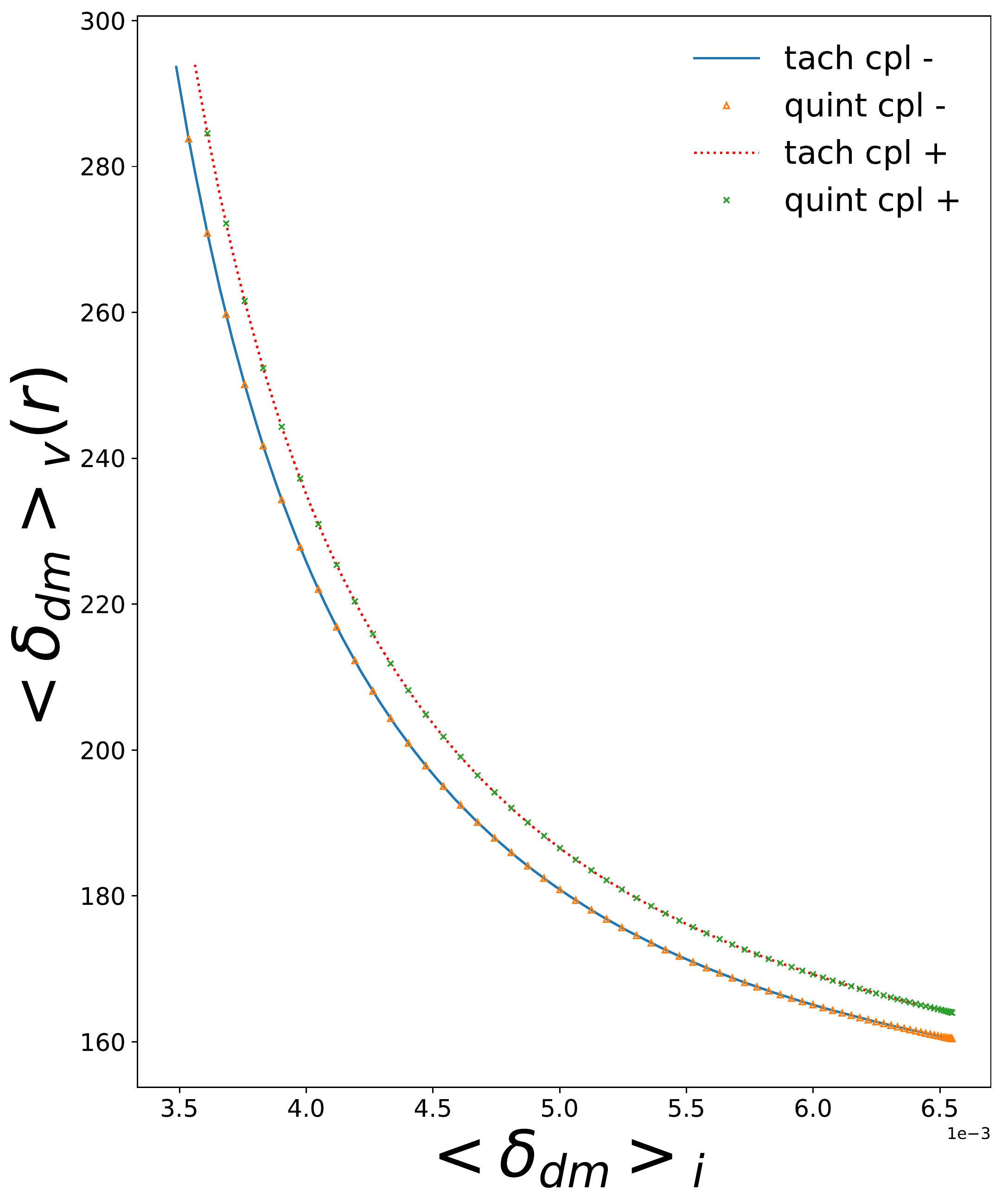}
\caption{\label{fig:6cpl} Virial characteristics for CPL case. Left
  panel shows ratio of virial radius to turn around radius as a
  function of the initial density contrast in dark matter. Right panel
  shows  Density contrast at virialisation as a function of the
  initial density contrast in dark matter. "quint" denotes quintessence
  and "tach" represents tachyonic field. cpl+ denotes $w_a = +0.09$ and cpl-
  represents  $w_a = -0.09$.}
\end{figure}

We present the characteristics of virialization in
Figure~\ref{fig:6cpl}.
We have plotted the ratio of the virial radius to the turn around
radius in the left panel as a function of the initial density
contrast.
The expected value for this ratio is $0.5$ in the Einstein-deSitter
model.
In case dark energy clusters significantly and also participates in
the virialization process, the expected value is above $0.5$, and if
dark energy clustering is not relevant to the virialization process
then the expected value is below $0.5$ \cite{Maor:2005hq}.
In the right panel we have plotted the density contrast at the time of
virialization.
Here, virialization is defined by the epoch at which 
\begin{equation}
<T> + \frac{1}{2}\left\langle R \, F_{R} \right\rangle = 0 \label{vireq:1}
\end{equation}
here $T$ is the kinetic energy, $R$ is the radius of the shell and
$F_R$ is the radial force on the shell, see \cite{2018JCAP...06..018P}
for details. 
Thus the volume averaged overdensity within a virialized shell is
expected to be around $145$ in Einstein-deSitter model.
In the $\Lambda$CDM model, the expected value is higher as
perturbations take a longer time to collapse.
Further, as we approach the critical density contrast for collapse
from above, the density contrast at virialization shoots up.
Similar behaviour is observed for quintessence models 
\cite{2018JCAP...06..018P,erratum}.
We see that the qualitative behaviour for the two CPL cases is similar
to that for $\Lambda$CDM and that seen for some quintessence models.
The ratio of virial radius to the turn around radius varies almost in
the same manner for the two CPL models with small differences for
large initial density contrast.
There are no systematic differences between tachyon and quintessence
models for a given CPL prescription for the equation of state
parameter.
Curves for the two CPL models differ clearly from each other in the
right panel but again, there are no differences between tachyon and
quintessence models for a given set of CPL parameters.

These results are remarkable in that it appears that we can ignore the
precise choice of dark energy model and use any convenient
prescription as long as we get the same expansion history.
This can be done if our interest is restricted to perturbations in
dark matter.

\subsection{Perturbations in Dark Energy}
\label{subsec:de_results}

\begin{figure}[hbt!]
\centering 
\includegraphics[width=0.8\textwidth]{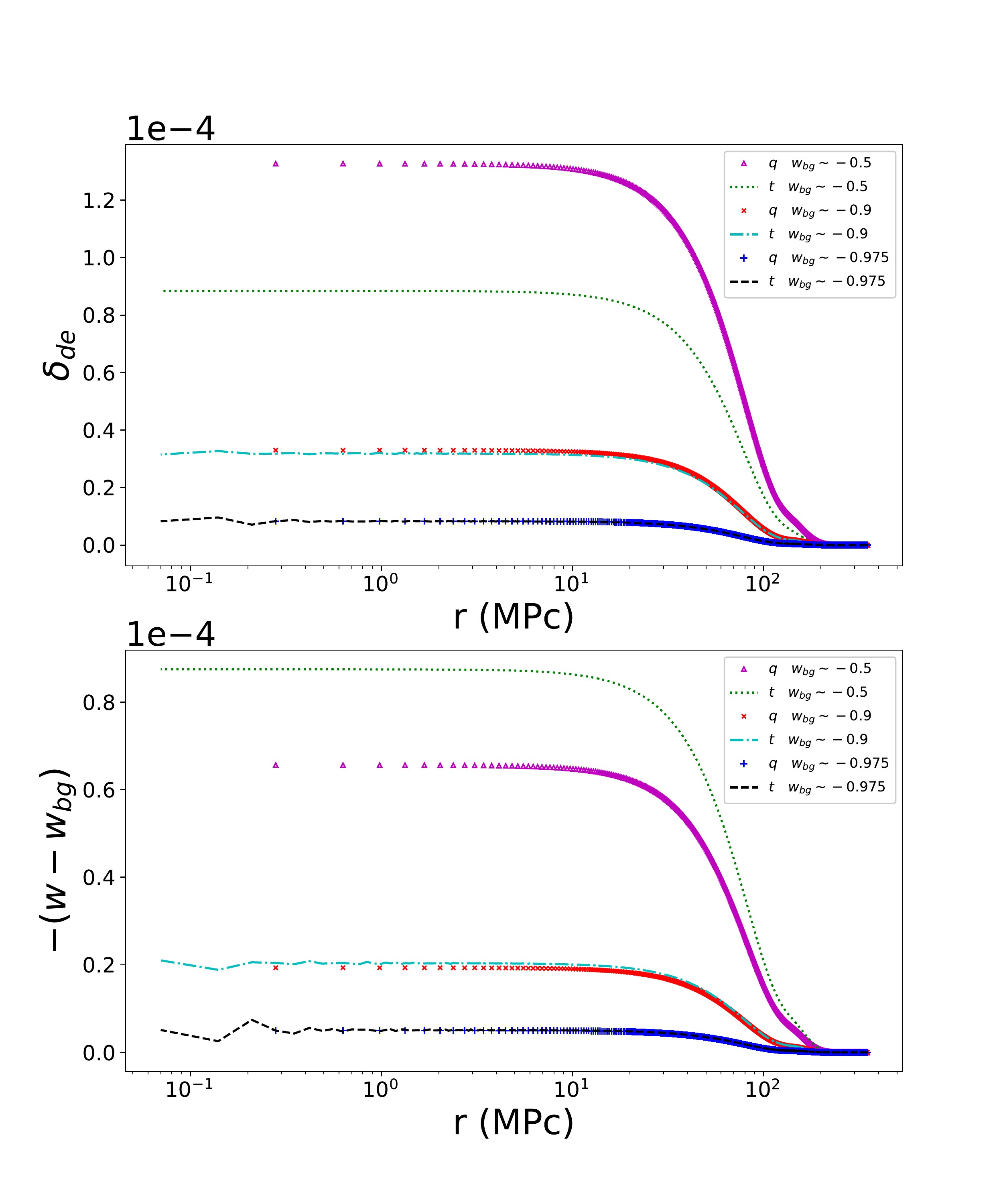}
\caption{\label{fig:9}  Underdense case:(Upper panel) Dark energy
  density contrast evolved to $z\sim 0$ . q denotes quintessence and t
  represent tachyonic field. $w_{bg}$ for constant value of background
  equation of state for dark energy field. This is for initially
  underdense case(UD1). Lower panel: Equation of state comparison for
  three constant equation of state cases.}     
\end{figure}

We now turn our attention to perturbations in dark energy. 
We study two physical quantities, density contrast for dark energy
$\delta_{de}$ and the equation of state parameter $w$. 
These are shown as a function of radius for an initially underdense
matter perturbation. 
We have plotted $\delta_{de}$ as a function of $r$ for constant $w$
cosmologies in Figure~\ref{fig:9}(upper panel).
Curves are plotted at $z=0$ and refer to the simulations used in
Figure~6. 
We see that density contrast in dark energy remains small at all
scales.
The amplitude of dark energy perturbations is higher when the model
deviates significantly from $\Lambda$CDM: we see that the amplitude is
highest for the model with $w=-0.5$ and decreases for models with a
smaller $w$.
We see that the curves for each $w$ are distinct.
We also note that the tachyon models and quintessence models differ
from each other and this difference is larger for models
with a larger $w$.
We have shown in earlier work that $w$ becomes a function of space for
dynamical dark energy models.
Variations from the expected value in the background for constant $w$
models is shown in Figure~\ref{fig:9} (lower panel) as a function of
$r$. 

We see that for an underdensity in matter, $w$ is smaller than the
value in the background model.
Deviations are larger for models that deviate significantly from the
$\Lambda$CDM models.
Differences between tachyon models and quintessence models can be seen
and these are larger for the models with a larger $w$.

\begin{figure}[hbt!]
\centering 
\includegraphics[width=0.8\textwidth]{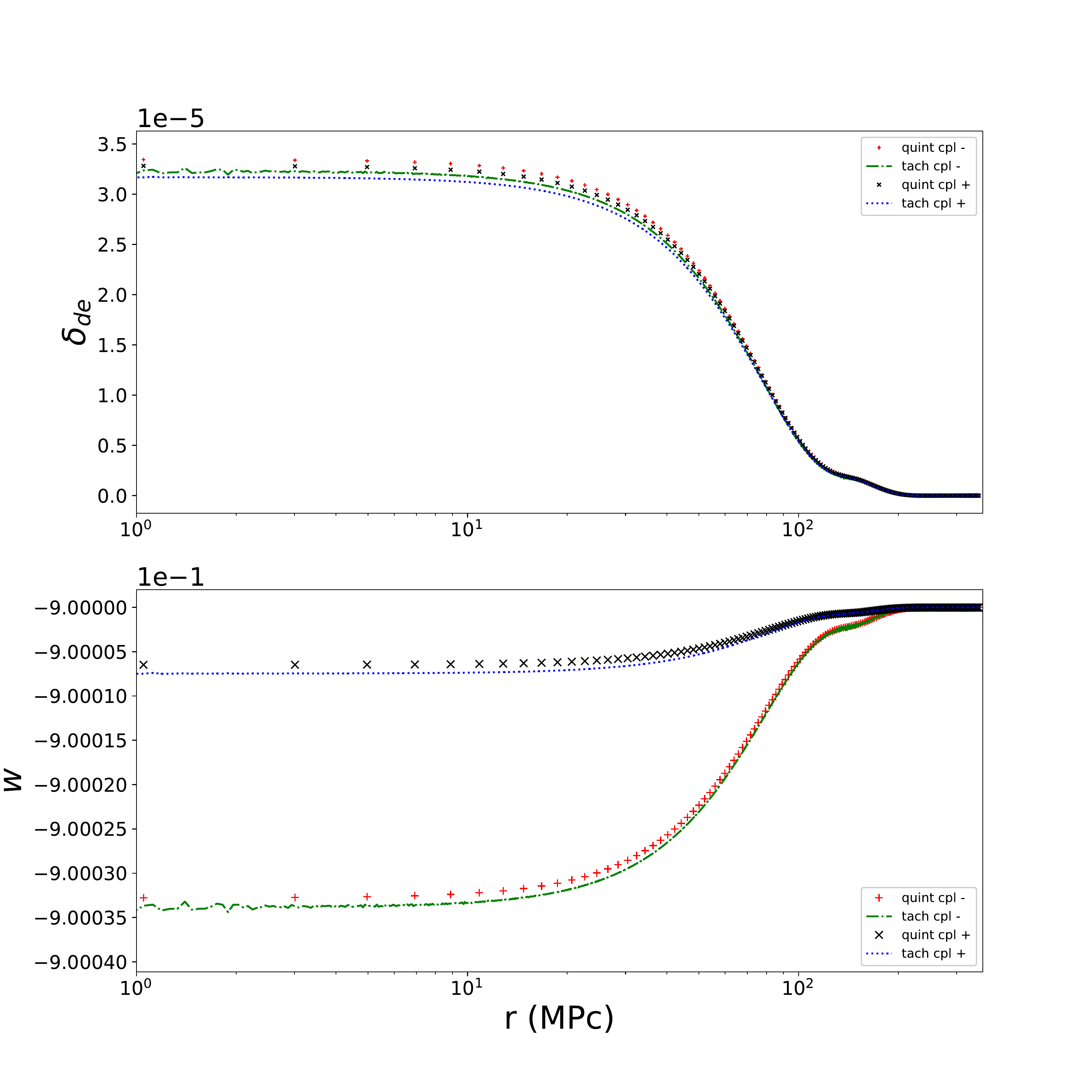}
\caption{\label{fig:9cpl}Underdense CPL case:(Upper Panel) Dark energy
  density contrast evolved to $z\sim 0$ . q denotes quintessence and t
  represent tachyonic field. cpl+ denotes $w_a = +0.09$ and cpl- represents
  $w_a = -0.09$. This for initially underdense case(UD1cpl). Lower
  Panel:  Equation of state($w$) evolved to $z\sim 0$.  This is for
  initially underdense case(UD1cpl).}  
\end{figure}

Plots for CPL models are given in Figure~\ref{fig:9cpl}.
Both the models here are consistent with most low redshift
observations \cite{Tripathi:2016slv}.
We keep $w_0=-0.9$ and $w_a=\pm 0.09$, we refer to these models as $cpl+$
or $cpl-$ depending on the sign of $w_a$.
These figures refer to the simulations used for Figure~7.
Quantities are plotted at $z=0$. 
We see that there are differences between the tachyon and quintessence
models for each CPL model but the differences remain small at all
scales. 

Unlike dark matter, we find that dark energy perturbations do carry an
imprint of the model.
Differences between tachyon and quintessence models for the same
expansion history become larger for models with large deviations from
the $\Lambda$CDM model.
Differences are small for constant $w$ models allowed by observations.

\section{Summary}
\label{sec:summary}

We have presented results of our study of evolution of perturbations
in dark matter and tachyon models of dark energy.
We find that differences across models arising from different
potentials are small.
As different potentials correspond to different expansion history, it
is difficult to dilineate the dependencies.

In order to study the dependence of evolution of perturbations on the
class of models, we construct potentials in quintessence and tachyon
models corresponding to constant equation of state $w$ for dark energy
and CPL parameterization.
This allows us to address the question of the dependence of evolution
of perturbations on the class of models.

We study spherically symmetric perturbations using a self-consistent
relativistic code. 
We study evolution of regions where dark matter is
underdense/overdense.

We find that evolution of dark matter perturbations depends only on
the expansion history.
There is no discernable imprint of the dark energy model on the
evolution of dark matter perturbations.

Dark energy perturbations remain small in all cases studied here.
The amplitude of dark energy perturbations depends on the expansion
history as well as the dark energy model (tachyon/quintessence).
\textit{Thus in principle there is an observable signature of the class of
dark energy models, though the differences are very small}.
These differences are larger for models that deviate significantly
from the $\Lambda$CDM model in terms of the expansion history.

  While the results follow from well defined theoretical
  models and numerical calculations, it is useful to have some physical
  insight.
  One can argue from continuity that as one goes towards the $\Lambda
  CDM$ limit of $w=-1$, all models should converge to $\Lambda$ like
  behaviour.
  One crucial point to check here is for the deviations of $w$ from
  $-1$, that are allowed by observations, can different models be
  distinguished by perturbations?
  In this article we have done nonlinear calculations to probe this
  question.
  One of the key takeaways from our work is that the two classes of
  models considered here are indistinguishable not only for cases very
  close to $\Lambda$ limit, but are so even for scenarios which are
  significantly different from $\Lambda$ limit.
  The above statement applies to characteristics of dark matter
  perturbations.
  We believe that this is due to matter being the dominant component
  for much of the expansion history and matter dominating over dark
  energy in regions with high overdensity of matter.
  While this has been pointed out in studies based on linear theory or
  heuristic arguments, we believe that the calculations presented in
  this manuscript establish this for the first time with self
  consistent and relativistic calculations in the non-linear regime.

The useful conclusion that we can draw from this study is that we may
choose any dark energy model to reproduce the appropriate expansion
history as the evolution of dark matter perturbations is insensitive
to the specifics of the dark energy model other than the expansion
history.

At the same time, the very small magnitude of differences of dark
energy perturbations indicate that it will be almost impossible for us
to discover the true dark energy model from measurements of distances
or characteristics of dark matter perturbations.

\ack

Authors acknowledge useful discussion with Harvinder K. Jassal and
Avinash Singh.
Computational work was carried out at computing facilities at IISER
Mohali. 
This research has made use of NASA's Astrophysics Data System
Bibliographic Services.


\section*{References}



\begin{thebibliography}{10}

\bibitem{Riess_1998}
A.~G. Riess, A.~V. Filippenko, P.~Challis, A.~Clocchiatti, A.~Diercks, P.~M.
  Garnavich, R.~L. Gilliland, C.~J. Hogan, S.~Jha, R.~P. Kirshner,
  B.~Leibundgut, M.~M. Phillips, D.~Reiss, B.~P. Schmidt, R.~A. Schommer, R.~C.
  Smith, J.~Spyromilio, C.~Stubbs, N.~B. Suntzeff, and J.~Tonry,
  ``Observational evidence from supernovae for an accelerating universe and a
  cosmological constant,'' {\em The Astronomical Journal}, vol.~116,
  pp.~1009--1038, sep 1998.

\bibitem{Perlmutter:1998np}
S.~Perlmutter {\em et~al.}, ``{Measurements of Omega and Lambda from 42 high
  redshift supernovae},'' {\em Astrophys. J.}, vol.~517, pp.~565--586, 1999.

\bibitem{1993ApJ...413L.105P}
M.~M. {Phillips}, ``{The absolute magnitudes of Type IA supernovae},'' {\em
  apjl}, vol.~413, pp.~L105--L108, Aug. 1993.

\bibitem{1996AJ....112.2391H}
M.~{Hamuy}, M.~M. {Phillips}, N.~B. {Suntzeff}, R.~A. {Schommer}, J.~{Maza},
  and R.~{Aviles}, ``{The Absolute Luminosities of the Calan/Tololo Type IA
  Supernovae},'' {\em aj}, vol.~112, p.~2391, Dec. 1996.

\bibitem{Riess_1996}
A.~G. Riess, W.~H. Press, and R.~P. Kirshner, ``A precise distance indicator:
  Type ia supernova multicolor light-curve shapes,'' {\em The Astrophysical
  Journal}, vol.~473, pp.~88--109, dec 1996.

\bibitem{1995ApJ...438L..17R}
A.~G. {Riess}, W.~H. {Press}, and R.~P. {Kirshner}, ``{Using Type IA supernova
  light curve shapes to measure the Hubble constant},'' {\em apjl}, vol.~438,
  pp.~L17--L20, Jan. 1995.

\bibitem{1995Natur.377..600O}
J.~P. {Ostriker} and P.~J. {Steinhardt}, ``{The observational case for a
  low-density Universe with a non-zero cosmological constant},'' {\em nat},
  vol.~377, pp.~600--602, Oct. 1995.

\bibitem{1996ComAp..18..275B}
J.~S. {Bagla}, T.~{Padmanabhan}, and J.~V. {Narlikar}, ``{Crisis in Cosmology:
  Observational Constraints on {\ensuremath{\omega}} and H $_{O}$},'' {\em
  Comments on Astrophysics}, vol.~18, p.~275, Jan 1996.

\bibitem{RevModPhys.61.1}
S.~Weinberg, ``The cosmological constant problem,'' {\em Rev. Mod. Phys.},
  vol.~61, pp.~1--23, Jan 1989.

\bibitem{Bull:2015stt}
P.~Bull {\em et~al.}, ``{Beyond $\Lambda$CDM: Problems, solutions, and the road
  ahead},'' {\em Phys. Dark Univ.}, vol.~12, pp.~56--99, 2016.

\bibitem{Navarro:1999fr}
J.~F. Navarro and M.~Steinmetz, ``{The core density of dark matter halos: a
  critical challenge to the lambda-cdm paradigm?},'' {\em Astrophys. J.},
  vol.~528, pp.~607--611, 2000.

\bibitem{DelPopolo:2016emo}
A.~Del~Popolo and M.~Le~Delliou, ``{Small scale problems of the $\Lambda$CDM
  model: a short review},'' {\em Galaxies}, vol.~5, no.~1, p.~17, 2017.

\bibitem{Clifton:2011jh}
T.~Clifton, P.~G. Ferreira, A.~Padilla, and C.~Skordis, ``{Modified Gravity and
  Cosmology},'' {\em Phys. Rept.}, vol.~513, pp.~1--189, 2012.

\bibitem{PhysRevLett.85.2236}
B.~Boisseau, G.~Esposito-Far\`ese, D.~Polarski, and A.~A. Starobinsky,
  ``Reconstruction of a scalar-tensor theory of gravity in an accelerating
  universe,'' {\em Phys. Rev. Lett.}, vol.~85, pp.~2236--2239, Sep 2000.

\bibitem{Fujii_2003}
Y.~Fujii and K.~ichi Maeda, ``The scalar-tensor theory of gravitation,'' {\em
  Classical and Quantum Gravity}, vol.~20, pp.~4503--4503, sep 2003.

\bibitem{Sotiriou:2008rp}
T.~P. Sotiriou and V.~Faraoni, ``{f(R) Theories Of Gravity},'' {\em Rev. Mod.
  Phys.}, vol.~82, pp.~451--497, 2010.

\bibitem{Starobinsky:2007hu}
A.~A. Starobinsky, ``{Disappearing cosmological constant in f(R) gravity},''
  {\em JETP Lett.}, vol.~86, pp.~157--163, 2007.
  
\bibitem{Nojiri:2010wj}
S.~Nojiri and S.~D.~Odintsov,
Phys.\ Rept.\  \textbf{505}, 59-144 (2011)
doi:10.1016/j.physrep.2011.04.001
[arXiv:1011.0544 [gr-qc]].


\bibitem{Nojiri:2017ncd}
S.~Nojiri, S.~Odintsov and V.~Oikonomou,
Phys.\ Rept.\  \textbf{692}, 1-104 (2017)
doi:10.1016/j.physrep.2017.06.001
[arXiv:1705.11098 [gr-qc]].

\bibitem{Will:2014kxa}
C.~M. Will, ``{The Confrontation between General Relativity and Experiment},''
  {\em Living Rev. Rel.}, vol.~17, p.~4, 2014.

\bibitem{Yunes2013}
N.~Yunes and X.~Siemens, ``Gravitational-wave tests of general relativity with
  ground-based detectors and pulsar-timing arrays,'' {\em Living Reviews in
  Relativity}, vol.~16, p.~9, Nov 2013.

\bibitem{PhysRevD.78.063503}
B.~Jain and P.~Zhang, ``Observational tests of modified gravity,'' {\em Phys.
  Rev. D}, vol.~78, p.~063503, Sep 2008.

\bibitem{2010deto.book.....A}
L.~{Amendola} and S.~{Tsujikawa}, {\em {Dark Energy}}.
\newblock Jan. 2015.

\bibitem{Buchert:2007ik}
T.~Buchert, ``{Dark Energy from Structure: A Status Report},'' {\em Gen. Rel.
  Grav.}, vol.~40, pp.~467--527, 2008.

\bibitem{Huterer:2017buf}
D.~Huterer and D.~L. Shafer, ``{Dark energy two decades after: Observables,
  probes, consistency tests},'' {\em Rept. Prog. Phys.}, vol.~81, no.~1,
  p.~016901, 2018.

\bibitem{Copeland:2006wr}
E.~J. Copeland, M.~Sami, and S.~Tsujikawa, ``{Dynamics of dark energy},'' {\em
  Int. J. Mod. Phys.}, vol.~D15, pp.~1753--1936, 2006.

\bibitem{Ratra:1987rm}
B.~Ratra and P.~J.~E. Peebles, ``{Cosmological Consequences of a Rolling
  Homogeneous Scalar Field},'' {\em Phys. Rev.}, vol.~D37, p.~3406, 1988.

\bibitem{PhysRevD.67.063504}
J.~S. Bagla, H.~K. Jassal, and T.~Padmanabhan, ``Cosmology with tachyon field
  as dark energy,'' {\em Phys. Rev. D}, vol.~67, p.~063504, Mar 2003.

\bibitem{ArmendarizPicon:2000dh}
C.~Armendariz-Picon, V.~F. Mukhanov, and P.~J. Steinhardt, ``{A Dynamical
  solution to the problem of a small cosmological constant and late time cosmic
  acceleration},'' {\em Phys. Rev. Lett.}, vol.~85, pp.~4438--4441, 2000.

\bibitem{ArmendarizPicon:2000ah}
C.~Armendariz-Picon, V.~F. Mukhanov, and P.~J. Steinhardt, ``{Essentials of k
  essence},'' {\em Phys. Rev.}, vol.~D63, p.~103510, 2001.

\bibitem{2001PhLB..511..265K}
A.~{Kamenshchik}, U.~{Moschella}, and V.~{Pasquier}, ``{An alternative to
  quintessence},'' {\em Physics Letters B}, vol.~511, pp.~265--268, Jul 2001.

\bibitem{Doran:2003xq}
M.~Doran, C.~M. Muller, G.~Schafer, and C.~Wetterich, ``{Gauge-invariant
  initial conditions and early time perturbations in quintessence universes},''
  {\em Phys. Rev.}, vol.~D68, p.~063505, 2003.

\bibitem{Malquarti:2002iu}
M.~Malquarti and A.~R. Liddle, ``{Evolution of large scale perturbations in
  quintessence models},'' {\em Phys. Rev.}, vol.~D66, p.~123506, 2002.

\bibitem{Abramo:2001mv}
L.~R.~W. Abramo and F.~Finelli, ``{Attractors and isocurvature perturbations in
  quintessence models},'' {\em Phys. Rev.}, vol.~D64, p.~083513, 2001.

\bibitem{Unnikrishnan:2008qe}
S.~Unnikrishnan, H.~K. Jassal, and T.~R. Seshadri, ``{Scalar Field Dark Energy
  Perturbations and their Scale Dependence},'' {\em Phys. Rev.}, vol.~D78,
  p.~123504, 2008.

\bibitem{Jassal:2009ya}
H.~K. Jassal, ``{A comparison of perturbations in fluid and scalar field models
  of dark energy},'' {\em Phys. Rev.}, vol.~D79, p.~127301, 2009.

\bibitem{PhysRevD.66.021301}
T.~Padmanabhan, ``Accelerated expansion of the universe driven by tachyonic
  matter,'' {\em Phys. Rev. D}, vol.~66, p.~021301, Jun 2002.
  
\bibitem{Sen:2002in}
A.~Sen,
JHEP \textbf{07} (2002), 065
doi:10.1088/1126-6708/2002/07/065
[arXiv:hep-th/0203265 [hep-th]].

  
\bibitem{Padmanabhan:2002sh}
T.~Padmanabhan and T.~R.~Choudhury,
Phys. Rev. D \textbf{66} (2002), 081301
doi:10.1103/PhysRevD.66.081301
[arXiv:hep-th/0205055 [hep-th]].


\bibitem{2019JCAP...04..047S}
  Avinash Singh, Archana Sangwan and H. K. Jassal, ``{Low redshift
    observational constraints on tachyon models of dark energy},''
  {\em jcap}, vol.~2019, p.~047, April 2019.
  
\bibitem{Singh:2019bfd}
A.~Singh, H.~K. Jassal, and M.~Sharma, ``{Perturbations in Tachyon Dark Energy
  and their Effect on Matter Clustering},'' 2019.

\bibitem{2018JCAP...06..018P}
M.~{Pratap Rajvanshi} and J.~S. {Bagla}, ``{Nonlinear spherical perturbations
  in quintessence models of dark energy},'' {\em jcap}, vol.~6, p.~018, June
2018.

\bibitem{erratum}
Manvendra~Pratap Rajvanshi and J.S Bagla.
\newblock Erratum: Nonlinear spherical perturbations in quintessence models of
  dark energy.
\newblock {\em Journal of Cosmology and Astroparticle Physics},
  2020(03):E01--E01, mar 2020.


\bibitem{Chevallier:2000qy}
M.~Chevallier and D.~Polarski, ``{Accelerating universes with scaling dark
  matter},'' {\em Int. J. Mod. Phys.}, vol.~D10, pp.~213--224, 2001.

\bibitem{Linder:2002et}
E.~V. Linder, ``{Exploring the expansion history of the universe},'' {\em Phys.
  Rev. Lett.}, vol.~90, p.~091301, 2003.

\bibitem{Rajvanshi:2019wmw}
M.~P.~Rajvanshi and J.~S.~Bagla,
J. Astrophys. Astron. \textbf{40} (2019) no.6, 44
doi:10.1007/s12036-019-9613-2
[arXiv:1905.01103 [astro-ph.CO]].

\bibitem{1993MNRAS.262..717B}
J.~D. {Barrow} and P.~{Saich}, ``{Growth of large-scale structure with a
  cosmological constant},'' {\em mnras}, vol.~262, pp.~717--725, Jun 1993.

\bibitem{Maor:2005hq}
I.~Maor and O.~Lahav, ``{On virialization with dark energy},'' {\em JCAP},
  vol.~0507, p.~003, 2005.

\bibitem{Tripathi:2016slv}
A.~Tripathi, A.~Sangwan, and H.~K. Jassal, ``{Dark energy equation of state
  parameter and its evolution at low redshift},'' {\em JCAP}, vol.~1706,
  no.~06, p.~012, 2017.

\end{thebibliography}
\end{document}